\documentclass[manuscript, nonacm]{acmart}

\acmBooktitle{2026 CHI Conference on Human Factors in Computing Systems (CHI '26), April 13--17, 2026, Barcelona, Spain}

\usepackage{blindtext}

\AtBeginDocument{%
  \providecommand\BibTeX{{%
    \normalfont B\kern-0.5em{\scshape i\kern-0.25em b}\kern-0.8em\TeX}}}

\begin{document}

\title{Robotic Affection - Opportunities of AI-based haptic interactions to improve social robotic touch through a multi-deep-learning approach}

\author{Ali Askari}
\affiliation{%
  \institution{TU Dortmund University}
  \city{Dortmund}
  \country{Germany}}
\email{ali.askari@tu-dortmund.de}
\orcid{0000-0002-4374-3635}

\author{Jens Gerken}
\affiliation{%
  \institution{TU Dortmund University}
  \city{Dortmund}
  \country{Germany}}
\email{jens.gerken@tu-dortmund.de}
\orcid{0000-0002-0634-3931}

\renewcommand{\shortauthors}{Askari \& Gerken}

\begin{abstract}
  Despite the advancement in robotic grasping and dexterity through haptic information, affective social touch, such as handshaking or reassuring stroking, remains a major challenge in Human-Robot-Interaction. This position paper examines current progress and limitations across artificial intelligence, haptics and robotics research, and proposes a novel multi-model architecture to address these gaps. Drawing inspiration from neurobiology, we decompose affective touch into distinct, specialized subtasks models. By treating affective touch as a distributed, closed-loop perceptual task rather than a monolithic motoric movement, we aim to overcome the "haptic uncanny valley" through a peer-to-peer, state-sharing framework. Our approach supports scalable and cumulative development within a Sim-to-Real pipeline, fostering interdisciplinary collaboration. By enabling haptics, AI, and robotics researchers to contribute independently yet coherently, we outline a pathway toward a unified, expressive system for social robotics. 
\end{abstract}



\maketitle

\section{Introduction}
In many global cultures, the foundational interaction of a social greeting is the handshake. Far from being a mere motoric movement, this act constitutes a rich channel of non-verbal communication that can convey emotions and establish social rapport between the involved parties \cite{hertenstein_touch_2002, hertenstein_touch_2006}. While these and other affective touch interactions are fundamental in human-to-human encounters \cite{krahe_meaning_2024}, particularly in the context of care and assistance, they remain a significant challenge in Human–Robot Interaction (HRI). Especially in fields like rehabilitation science where robots are used in the context of care or assistance, such affective haptic interactions could carry the potential of improving the acceptance of robots and well-being of patients.\\
While general robotic movements, context-awareness, speech and also the usage of haptic processes for grasping in robotic interactions has improved significantly due to advancements in several fields of artificial intelligence (AI) \cite{huang_vt-refine_2025, grandia_design_2024}, there is still a gap between advancements in haptic design and its transfer to \textit{affective} robotic interactions. Those still lack the fine-tuned haptic stimulation that makes for example a hand-shake "affective" and human-like. One contributing factor for why this interaction fails is the nature of handshaking and other "affective haptic interactions", such as a reassuring pat or empathic stroking. These interactions rely on the covariance of skin deformation and joint angles to specify the "meaning" of touch. During such contact, humans perceive both normal and shear forces. The human is able to distinguish, interpret, and react to these forces \cite{dargahi_human_2004}. As Gibson describes it, social touch is an active perceptual system rather than a passive sensation \cite{gibson_senses_1969}.  For that, a system must move beyond the mere replication of human kinematic trajectories. It must instead function as an active perceptual system capable of adjusting its output based on the re-afferent feedback obtained from the human recipient. Early models like Adams "Closed-Loop-Theory" \cite{adams_closed-loop_1971} emphasize the role of a perceptual trace in human motor-control, and therefore a truly human-centered robotic system requires the real-time integration of exactly this re-afferent feedback. This can only be achieved by interdisciplinary work of haptic, robotics and AI research. 

In this position paper, we analyze the topics and trends that could open up opportunities to improve affective haptic interactions and argue that considering a closed-haptic feedback loop in the design of hand-based human-robot interactions is crucial to overcome this type of "haptic uncanny valley". We describe a possible pathway of achieving this by combining haptics as its own neural model in a peer-to-peer architecture with other system models, utilizing advancements in the fields of Behavioral Cloning and Reinforcement Learning. Furthermore this approach could benefit haptic, AI and robotic researchers by creating a distributed Sim-to-Real pipeline where all three fields could contribute independently to their areas of expertise. 

We will first examine the embodied AI models involved, before looking into the data acquisition required to feed the learning processes of the models in the first place. This way, by understanding how a model learns to resonate to the haptic stimulation patterns of a social greeting, we can better understand the specific requirements for the haptic information that must be captured during the demonstration phase. 

\section{Reinforcement Learning from Demonstration}
GPU-based physics simulations, such as NVIDIA Isaac Gym \cite{makoviychuk_isaac_2021}, have emerged as powerful environments for the development of \textit{Reinforcement Learning} (RL) policies. By leveraging massive parallelism to simulate thousands of environments simultaneously, these systems enable AI agents to accumulate years of physical experience within minutes. They provide high-fidelity modeling of rigid body dynamics, collision detection, and joint constraints, allowing for accurate simulation of complex physical interactions. The resulting large-scale datasets support the iterative refinement of control policies, which can subsequently be deployed on physical hardware through Sim-to-Real transfer. \\
In the domain of robotic hands, RL has enabled substantial advances in both locomotion \cite{sasaki_locomotion_2025} and dexterous manipulation \cite{li_grasp_2024, wang_anthropomorphic_2024, yu_dexterous_2022}. These works demonstrate that high-dimensional control of anthropomorphic systems can be achieved with increasing robustness and adaptability. Collectively, they establish a strong methodological foundation for extending learned control paradigms beyond purely functional manipulation toward more expressive interaction modalities.\\
Within this framework, Deep Reinforcement Learning (DRL) enables agents to optimize behavior by maximizing cumulative reward signals. Reward signals are typically formulated as a weighted sum of multiple sub-objectives. The agent's goal is to maximize the expected cumulative reward over time, usually going through a Markov Decision Process \cite{santana_introduction_2025}. These approaches have proven particularly effective in domains characterized by non-linear dynamics. In animatronics, DRL-based systems synthesize lifelike, robust movements that generalize across diverse interaction contexts and environmental conditions \cite{grandia_design_2024}. The capacity of DRL agents to autonomously discover unanticipated strategies makes them especially well-suited for complex control tasks such as vehicle guidance and legged locomotion. However, one of the more work-intensive tasks of this approach is the manual parameter fine-tuning of the aforementioned reward functions, often referred to as \textit{reward engineering}. For complex interactions, designing those hand-crafted reward functions become increasingly difficult to achieve using conventional models. 

\section{Behavioral Cloning}
An alternative yet complementary paradigm is supervised learning through \textit{Behavioral Cloning} (BC). Instead of relying on agent-generated simulation data and pre-defined reward functions, BC directly learns policies from expert demonstrations of recorded datasets or teleoperation data \cite{hoque_egodex_2025, oneill_open_2024, goldau_dormadl_2023}. The objective of the agent is to directly replicate the behavior observed in those large-scale datasets acquired prior to training. By utilizing policies optimized through loss functions rather than weighted reward objectives like in DRL models \cite{santana_introduction_2025}, BC circumvents the need for extensive manual fine-tuning. Using stochastic instead of deterministic policies, diffusion-based BC enables much more refined sequence of actions \cite{chi_diffusion_2025}. This data driven natures makes it particularly attractive for capturing fluid, non-linear behaviors that are difficult to specify analytically.\\
Diffusion-Based BC has already been successfully integrated with tactile sensing pipelines in robotic hands \cite{zhang_unitachand_2025, huang_vt-refine_2025}, demonstrating its capacity to model complex haptic feedback in manipulation tasks. A challenge of standard BC without diffusion-based policies is the issue of covariate shift, where small execution errors compound over time, can cause policies to drift into unseen states \cite{sugiyama_covariate_2007}. Solutions to target this issue by using mitigation strategies such as dataset aggregation and regularization have been proposed and are currently used \cite{laskey_dart_2017, popov_mitigating_2024}.

\section{Data collection of hand interaction}

The acquisition of expert data for BC has likewise seen significant advancements. Teleoperation systems, motion-capture, and oracle policy generation pipelines provide structured mechanisms for collecting high-dimensional demonstrations \cite{wu_gello_2024}. In dexterous robotic hands applications, motion-capture gloves combined with inertial tracking have enabled increasingly natural demonstrations \cite{qin_anyteleop_2024, wang_dexcap_2024, fang_robotic_2015}. Recent research further integrates tactile feedback into this data-frameworks \cite{buamanee_bi-act_2024, huang_3d-vitac_2025, xu_dexumi_2025, huang_vt-refine_2025}, marking an important step toward multi-modal learning of contact-rich interactions.\\
Simultaneously, engineering advancements in sensing technology continue to improve spatial resolution, enabling multi-modal perception, and real-time integration \cite{hardman_multimodal_2025, song_haptic_2025}. Vision-based tactile sensors and high-resolution force sensing technologies are progressively enhancing the fidelity with which contact dynamics can be recorded and reproduced, although still difficult to embed in a wearable format. Together, these developments indicate steady progress toward richer, more expressive forms of robotic haptic interaction.

\section{Current challenges for affective haptic interaction}
Despite these substantial advancements, several challenges remain when extending current methodologies to affective social touch.\\
First, in RL, reward formulation remains a central bottleneck. Complex control environments typically require reward functions defined as weighted sums of multiple sub-objectives. While structured reward design has enabled impressive functional behaviors, fine-tuning these parameters with methods of reward engineering remains labor-intensive, even though work in this field aims to tackle this issue \cite{biyik_learning_2022}. Within the context of nuanced emotional interactions, the qualitative and subjective nature of the interaction makes the design of an effective, hand-crafted reward function increasingly difficult to achieve using conventional methods.\\
Second, Behavioral Cloning as previously mentioned faces stability challenges. While those are being addressed and not so critical in diffusion-based BC, a more fundamental bottleneck is the acquisition of the dataset itself. Collecting high-quality, high-dimensional expert data remains both resource-intensive and time-consuming. Robustness in long-horizon, contact-rich interactions remains therefore difficult to guarantee.\\
Finally, data acquisition for affective haptic interaction presents unresolved engineering constraints. Although, as mentioned, tactile integration is improving, existing pipelines primarily focus on functional grasping and manipulation \cite{buamanee_bi-act_2024, huang_3d-vitac_2025, xu_dexumi_2025, huang_vt-refine_2025}. Dedicated data-collection frameworks explicitly targeting affective hand interactions are, as far as we know, largely absent. A key obstacle is the trade-off between tactile resolution and ergonomic wearability. High-resolution sensing, particularly for shear forces essential to expressive touch, often relies on vision-based systems that are difficult to embed in wearable format \cite{choi_deep_2022, huang_vt-refine_2025}. Additionally, tactile morphological alignment between human demonstrators and robotic hands remains challenging \cite{zhang_unitachand_2025}, due to significant size and shape discrepancies between a human experts hand and the robotic sensor layer which complicate the transfer of these delicate affective patterns.\\
In the light of these challenges, we propose the following multi-model approach, aiming at addressing these limitations and enabling affective haptic interaction in robotic systems. 

\section{Multi-Model approach for social robotics}
\begin{figure}
  \includegraphics[width=\textwidth]{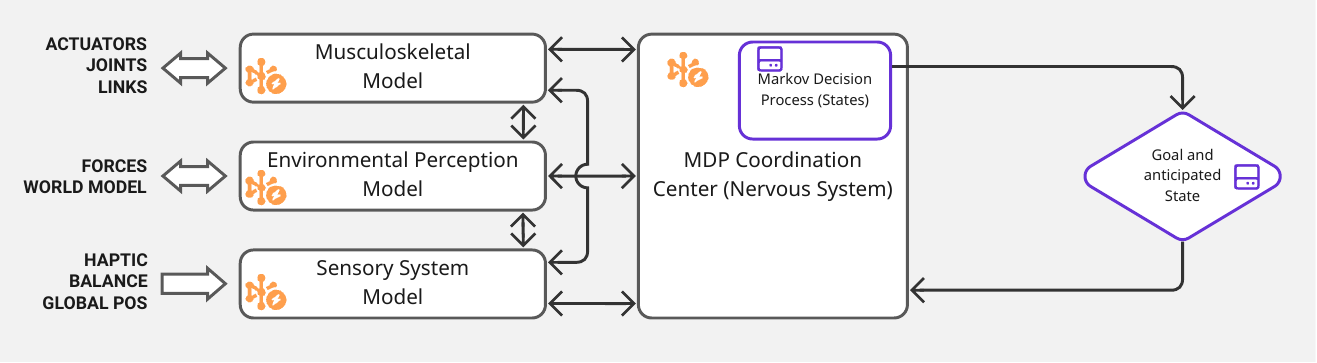}
  \caption{Architecture of a possible adaption from Bernsteins System Motor Control Theory \cite{bernstein_co-ordination_1967} to a robotic control architecture by dividing the system into four neural models.}
  \Description{Overview of the described architecture. The musculosketal, environmental, sensory and nervous system are represented as models and interconnected. The musculosketal is getting and sending information to and from actuators, joints, and links. The environmental is giving and receiving information from forces and the world model. The sensory is receiving information from haptic, balance, and global position. The nervous systems markov decision process state is calculating goal and anticipated state, which is giving back to the nervous system and from there on shared.}
  \label{fig:modeloverview}
\end{figure}
Drawing inspiration from neurobiology, we suggest that affective touch and similar complex interactions should not be treated as monolithic tasks for a single neural model. Instead, these interactions are better modeled as a sum of distinct sub-processes, each managed by a specialized embodied AI model optimized for that specific sub-task. However, the partition and orchestration of these sub-tasks remains a core challenge, as there is currently no universally accepted neuro-biological theory on how neural systems control (automatic) movements \cite{cano-de-la-cuerda_theories_2015}. \\
One popular theory among neuro-biologist is Bernstein's System Theory of motor control \cite{bernstein_co-ordination_1967}. Bernstein defines movement as a self-organizing functional system emerging from the interaction of the \textit{nervous system}, the \textit{musculoskeletal system}, the \textit{sensory system}, and \textit{environmental forces} \cite{bernstein_co-ordination_1967, cano-de-la-cuerda_theories_2015}. We consider a simplified version of this theory to be appropriate for adapting it into in neural-model architecture. By integrating this theory with a Markov Decision Process (MDP), we propose a peer-to-peer embodied AI network as shown in figure \ref{fig:modeloverview} consisting of four distinct neural models 
\begin{itemize}
    \item The Musculoskeletal Model: Coordinates joints and actuators to produce force and move segments, while also physically constraining motions
    \item The Environmental Perception Model: Collects and interprets external forces and environmental context, while holding a predictive model of the real world
    \item The Sensory System Model: Detects the environment and processes haptic stimulation patterns, balance and skeletal position 
    \item The MDP Coordination Center: Serves as the "nervous system", which is in Bernstein's theory responsible for goal-setting and overall coordination of the systems. Distributes actions for the next anticipated state and plans trajectories
\end{itemize}
While the precise design of the architecture, neural model and especially the mathematical functions involved in those require further research, our general approach proposes that each system shares its internal state enriched with meta-data regarding current priorities and goal-influencing information. This data provides a comprehensive state for the MDP, which determines the next optimal state and triggers an action for the transition, which is shared across the various subsystems. 

\section{Conclusion} 
The proposed approach is conceptual at this stage and focuses on the model problem, not the data acquisition problem. We believe that by decomposing the overall system into distinct sub-tasks, each governed by a dedicated model, this approach has the potential to provide a significant step forward to enable the development of affective robot touch interactions.\\
Achieving this task could certainly also be implemented through multiple other strategies that include the integration of haptic data, in particular sensory input from regions such as the dorsal hand and the thenar web-space, both of which play an important role in social touch interaction like handshakes. We argue that our concept of a framework with distinct models would especially promote interdisciplinary collaboration among researchers in haptics, robotics and artificial intelligence. Establishing a shared orchestration mechanism and standardized state-sharing protocol would enable each research community to develop and share models and curate datasets aligned with its domain-specific expertise. These datasets could subsequently be reused within the same field to refine existing models or adopted across disciplines to support the development of complementary modeling approaches. Moreover, training and evaluation could be conducted within adoptable simulated environments, allowing systematic testing and benchmarking across diverse interaction scenarios. Such a structured and collaborative framework has the potential to support sustainable and cumulative progress in the field.  
\bibliographystyle{ACM-Reference-Format}
\bibliography{bibliography}

\section*{Authors}
\noindentparagraph{\href{https://ihri.reha.tu-dortmund.de/team/ali-askari/}{Ali Askari}} is a research associate and PhD student in the Inclusive HRI group at TU Dortmund University. His research focuses on shared control, robotic teleoperation, and hybrid work environments. He also teaches courses in rapid prototyping and has a strong interest in topics related to the maker community.. 

\noindentparagraph{\href{https://ihri.reha.tu-dortmund.de/research-unit/team/jens-gerken/}{Jens Gerken}} is a Full Professor of Inclusive HRI at TU Dortmund University. His research covers gesture-based interaction design \cite{FOEHRENBACH2009, Jetter2010_Natural}, post-wimp and tangible interfaces \cite{jetter2011, Jetter2012}, shared control and teleoperation \cite{pascher2024humanai, Arevalo2021}. He also investigates alternative input/output modalities and sensors in the context of HRI and human-technology interaction~\cite{Woehle2018}, with a particular focus on accessibility and the needs of people with disabilities.

\end{document}